# Another Angle on Benchmarking Noncovalent Interactions

Vladimir Fishman, Michał Lesiuk, Jan M. L. Martin,* and A. Daniel Boese*

**ABSTRACT:** For noncovalent interactions, the CCSD(T)-coupled cluster method is widely regarded as the "gold standard". With localized orbital approximations, benchmarks for ever larger complexes are being published, yet FN-DMC (fixed-node quantum Monte Carlo) intermolecular interaction energies diverge to a progressively larger degree from CCSD(T) as the system size grows, particularly when $\pi$-stacking is involved. Unfortunately, post-CCSD(T) methods like CCSDT(Q) are cost-prohibitive, which requires us to consider alternative means of estimating post-CCSD(T) contributions. In this work, we take a step back by considering the evolution of the correlation energy with respect to the number of subunits for such $\pi$-stacked sequences as acene dimers and alkadiene dimers. We show it to be almost perfectly linear and propose the slope of the line as a probe for the behavior of a given electron correlation method. By going further into the coupled cluster expansion and comparing with CCSDT(Q) results for benzene and naphthalene dimers, we show that CCSD(T) does slightly overbind but not as strongly as suggested by the FN-DMC results.

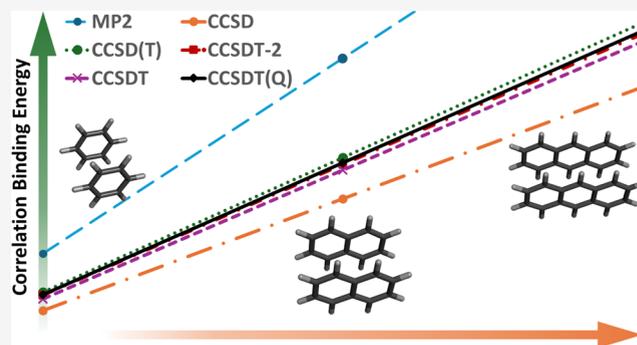

## 1. INTRODUCTION

Theoretical predictions have become an invaluable tool in modern science. As accurate benchmarks in quantum chemistry become more and more common and numerous,[1,2] some questions have arisen about the applicability of the underlying benchmark methods. For increasingly large systems,[3] there are growing discrepancies[4] between the interaction energies of large molecules for the main benchmark method of quantum chemistry, which is CCSD(T) (coupled cluster[5] including all single and double substitutions and quasiperturbative triple excitations[6]) and the main benchmark method of solid state materials science, which is QMC (quantum Monte Carlo).[7−10]

We are trying to glance at a piece of this puzzle by extrapolating dimer interactions to progressively larger scales using different methods. We will show that by investigating the slope of the intermolecular correlation energies with respect to monomer size, we can estimate the error of benchmark methods for much larger systems even than those calculated, gauging their accuracy.

Here, we will focus on CCSD(T) as the gold standard of quantum chemistry and examine all conceivable causes of its possible shortcomings:

1. Monomer polarizabilities will increase with the system size of aromatic molecules, as will the strength of their van der Waals interactions. This polarizability may thus become increasingly hard to calculate, requiring progressively larger basis sets. For instance, even small molecules and atoms with large polarizabilities require many diffuse functions in order to accurately describe these quantities.[11]

2. For calculating larger systems with CCSD(T), local orbital approximations are utilized. As the number of very small contributions to the interaction energy grows exponentially, their collective neglect might conceivably result in substantial errors for larger interacting systems. ("Many a little makes a mickle", as the English proverb goes.)

3. Larger intermolecular interactions may possibly be overestimated due to the quasiperturbative triples in the (T) part, as second-order Møller−Plesset perturbation theory (MP2) is known to strongly overestimate[12] intermolecular interactions of aromatic molecules. The cause of the failure of MP2 for intermolecular interactions of large systems has been rationalized by the incomplete "electrodynamic" screening of the Coulomb interaction.[13,14] Various papers have put forward this argument, showing that post-CCSD(T) contributions should become relevant at some point.[13−15] However, one has to bear in mind that, while MP2 overestimates interaction energies of $\pi$ stacks when comparing to more accurate reference methods, MP3 overcorrects, and hence underbinds them (which has inspired the MP2.5 method[16]). CCSD can be seen as a summation of MPn singles, doubles, and their disconnected products to infinite order, and underestimates interaction energies as it neglects connected triples (T), which first appear at fourth order. In contrast, however, other post-Hartree−Fock or post-DFT methods such as RPA (random phase approximation[17]) and DFT-SAPT (density functional theory-based symmetry-adapted perturbation theory[18]) are argued to yield better intermolecular interaction energies for larger dimers.[13,14]

Somewhat related to this list is the effect coming from different geometries: While accurate CCSD(T) energies are comparatively easy to compute, CCSD(T) geometries are much scarcer, and few studies actually consider optimized intermolecular interaction geometries. This effect is mostly neglected, as fixed geometries are taken into account. For training or benchmarking lower-level methods like density functional theory (DFT), this a rational, or at least expected choice. Neglect of the 'relaxation energy', however, may conceivably affect comparisons with more accurate methods.

Fortunately, all these error sources can be tracked down and scrutinized for smaller systems:

1. For intermolecular interactions with medium to large basis sets, the exact basis set limit is usually bracketed by calculations excluding and including counterpoise corrections,[19,20] with very few notable exceptions, like, for example, the formic acid dimer.[21] Thus, an increasingly large gap between these two calculations (including and excluding a basis set correction) would indicate volatility of a certain post-Hartree−Fock method. Basis set incompleteness errors can often also be estimated using localized orbital or other reduced-scaling approaches. Another approach to mitigate basis set superposition error would be the use of plane wave basis sets.[22]

2. Unlike local methods, canonical post-Hartree−Fock methods do not neglect any orbitals when calculating interaction energies.

3. Post-Hartree−Fock methods such as coupled cluster theory[5] have a clear hierarchy. For example, when static correlation effects are present and not well-captured by CCSD(T), one may attempt to walk up the cluster expansion staircase to CCSDT and CCSDT(Q), which will yield some insight about whether "the gold standard" is really that golden.[23−26]

Concerning the effects of the geometries, while the use of CCSD(T) gradients is very often computationally prohibitive, several structures can be considered to obtain a more complete picture of the interaction. This is usually done by just varying the distances, e.g., extending the S66 set[27] of intermolecular interactions to the S66 × 8 set,[28] which considers each complex at eight different multiples of the *inter*monomer separation (with fixed *intra*monomer geometries), for a total of 528 data points.

By addressing all of these points above, we are capable of evaluating the accuracy of different methods, including the gold standard CCSD(T), for large systems.

The angle we propose is to compare the behavior of different correlation methods along sequences of successively expanded monomers. We shall show that, not only is this behavior rather linear, but that the slopes of the lines — specifically their deviation from the most accurate method — offer a new insight into the (mis)behavior of more approximate correlation methods.

## 2. COMPUTATIONAL DETAILS

The electronic structure calculations have been performed through various program systems, depending on the method employed. The DFT[29,30] geometry optimizations utilizing the $\omega$-B97M-V[31] functional as well as the DLPNO (Domain-Based Local Pair Natural Orbital)-CCSD(T)[32,33] calculations have been performed using ORCA[34] 5.0.3, DFT-SAPT[35,36] using MOLPRO[37] 2022.2 and PNO-LCCSD(T)[38] with domain approximations by means of MOLPRO[37] 2024.3, and MP3[39] using QChem[40] 5.4, exploiting its RI implementation.[41] All dRPA,[17,42] LNO (Local Natural Orbitals)-CCSD(T),[43,44] LNO−CCSDT(Q),[45] and canonical CCSD(T)[6,46] and CCSDT(Q)[47] calculations were carried out by means of the MRCC[48] (August 2023) program suite. All LNO−CCSD(T) calculations have been using their respective MP2 basis set corrections as implemented in the MRCC program. The LNO-cutoff for normal settings in this program is $10^{-5}$ $E_h$ for the occupied and $10^{-6}$ $E_h$ for the virtual space, for tight settings $3 \times 10^{-6}$ and $3 \times 10^{-7}$ $E_h$, for very tight settings $10^{-6}$ and $10^{-7}$ $E_h$, and for very very tight settings $3 \times 10^{-7}$ and $3 \times 10^{-8}$ $E_h$. Details of the algorithm can be found in ref 43. The PNO cutoff parameters for the `Tight` setting are as follows: PNO selection threshold for LMP2 based on natural occupation numbers is $10^{-8}$ $E_h$, PNO selection threshold for LCCSD based on occupation numbers is $10^{-8}$ $E_h$, occupation number threshold for selecting triples domains for the iterative (T) approximation is $10^{-7}$ $E_h$. In contrast, for the `vTight` setting, they are PNO selection threshold for LMP2 based on natural occupation numbers is $10^{-9}$ $E_h$, PNO selection threshold for LCCSD based on occupation numbers is $2.5 \times 10^{-9}$ $E_h$, occupation number threshold for selecting triples domains for the iterative (T) approximation is $5 \times 10^{-8}$ $E_h$. Details of the PNO domain approximations algorithm can be found in ref 38. While some canonical post-CCSD(T) calculations were also carried out using MRCC, the most demanding ones were performed using a development version of the CFOUR program system.[49] The DFT-SAPT calculations have been performed using the asymptotically corrected PBE0 functional,[50−53] where the monomer ionization potentials and energies of the highest occupied molecular orbital have been computed via PBE0 calculations using a cc-pVQZ basis set.

For the L7 (i.e., seven large noncovalent complexes) set of systems[54] and the buckycatcher,[4] geometries from the original references have been utilized as-is. All other structures were optimized with the $\omega$-B97M-V functional[31] using the QZVPP[55] basis set with full symmetry. For example, in the case of the benzene dimer, the $D_{6h}$ symmetric saddle point has been utilized. This implies that all structures are directly on top of each other, as only their interaction energy matters. They are not parallel-displaced, such as the minimum-energy structures of the benzene[56] and coronene dimers[57] — they are not minima at all. The monomer distances have either been fixed, staying within their respective distance for the polyene stacks, or completely relaxed within their symmetry for both the polyene as well as for the acene stacks. The rationale for this approach is to be able to use as much symmetry as possible, since only the slope of interaction energies of increasingly large dimers matters, and not their absolute interaction energy.

Table 1. Results of Different Methods for the Benzene Dimer, the Four Largest Intermolecular Complexes of the L7 Set, and the Buckycatcher[a,b]

|  | FN-DMC | MP3 | CCSD | DFT-SAPT | CCSD(T) | MP2.5 | MP2 |
|---|---|---|---|---|---|---|---|
| Benzene$_2$ PD | 10.0 ± 0.5,[4] | 7.2[27] | 6.1[67] | 11.8[68] | 11.2,[4] | 13.4[27] | 19.7,[27] 19.7[67] |
| GCGC | 52.0 ± 3.3,[4] 44.4 ± 2.5[69] | 36.0[54] | 35.7[67] | 54.7 | 57.0,[4] | 56.1[54] | 74.7,[54] 73.6,[70] 74.9[67] |
| C3A | 62.4 ± 4.2,[4] 69.5 ± 3.8[69] | 34.1[54] | 47.8[67] | 62.7 | 69.1,[4] | 74.7[54] | 113.0,[54] 109.2,[70] 109.7[67] |
| C2C2PD | 75.6 ± 3.3,[4] 73.2 ± 2.9[69] | 27.7[54] | 58.2[67] | 81.5 | 86.1,[4] | 95.4[54] | 159.9,[54] 154.3,[70] 155.7[67] |
| C3GC | 101.3 ± 5.4,[4] 105.0 ± 3.7[69] | 61.8[54] | 80.8[67] | 104.6 | 120.0,[4] | 127.2[54] | 188.8,[54] 182.2,[70] 183.7[67] |
| C60CPPA | 130.5 ± 5.9,[4] | | 93.8[67] | 157.8 | 174.6,[4] | | 354.9[67] |

[a]All values are reported in kJ/mol. [b]GCGC: guanine-cytosine tetramer; C3A: adenine··· circumcoronene; C2C2PD: coronene dimer; C3GC: circumcoronene··· guanine-cytosine; C60CPPA: buckminsterfullerene··· buckycatcher.

Throughout this study, Dunning's correlation consistent basis sets with and without diffuse functions, denoted cc-pV$n$Z[58] and aug-cc-pV$n$Z,[59] have been put to use and will sometimes be abbreviated by DZ, TZ, QZ, and 5Z, as well as aTZ, aQZ, and a5Z. The notations {T,Q}Z and {Q,5}Z refer to basis set extrapolations from TZ and QZ or QZ and 5Z basis sets according to the familiar $L^{-3}$ extrapolation formula, respectively.[60]

Unless specifically indicated otherwise, interaction energies have been counterpoise (CP) corrected by the standard Boys and Bernardi procedure.[61]

In order to shed additional light on the performance of CCSD(T)[6] vs fully iterative CCSDT[62] and CCSDT(Q),[47] — which have CPU time scalings with system size $O(N^7)$, $O(N^8)$, and $O(N^9)$, respectively — we considered one additional $O(N^7)$ post-CCSD(T) method, namely CCSDT-2.[62-64] As detailed by Cremer and co-workers,[65] CCSDT-2 omits all the $\hat{T}_3$ coupling terms from the full CCSDT $\hat{T}_3$ amplitudes equations (thus eliminating the $O(N^8)$ step) as well as the $\hat{T}_1$ coupling terms.[66]

$$E[\text{CCSDT}] - E[\text{CCSDT-2}] = E_{TT}^{[5]} + O(\lambda^6) \quad (1)$$

where $\lambda$ is the perturbation parameter and $E_{TT}^{[5]}$ is the fifth-order triples−triples interaction term. The additional terms CCSDT-2 carries beyond CCSD(T) again start out at fifth order:

$$E[\text{CCSDT-2}] - E[\text{CCSD(T)}] = E_{TQ}^{[5]} + O(\lambda^6) \quad (2)$$

where $E_{TQ}^{[5]}$ is the fifth-order triples-(disconnected) quadruples interaction term. CCSDT(Q) is exact to fifth order: the additional terms it introduces beyond CCSDT are

$$E[\text{CCSDT(Q)}] - E[\text{CCSDT}] = E_{QQ}^{[5]} + E_{QT}^{[5]} + O(\lambda^6) \quad (3)$$

where it should be noted[65] that $E_{QT}^{[5]}$ is the Hermitian conjugate of $E_{TQ}^{[5]}$.

## 3. RESULTS AND DISCUSSION

**3.1. L7 Set and Discrepancies between Fixed-Node Diffusion Monte Carlo and CCSD(T).** The original starting point of this study was, in fact, the discrepancy between the two reference methods CCSD(T) and FN-DMC (fixed-node diffusion Monte Carlo). For this purpose, it is useful to reiterate the differences between them. In addition, we have also performed DFT-SAPT calculations for all these systems. The theoretical background sections of refs 13,14. argue that DFT-SAPT may be even more accurate than CCSD(T) for the interaction energies of progressively larger systems. In Table 1, we show the DFT-SAPT together with the literature values of various post-Hartree−Fock methods and FN-DMC for four molecules of the L7 set[54] as well as the buckycatcher.

It is apparent from these numbers that MP2 often overestimates the interaction energies of large conjugated molecules, whereas MP3 underestimates them. MP2.5,[71] which relies on compensation between the errors of MP2 and MP3, still somewhat overestimates the interaction energies when going to larger molecules.[72] The values obtained by DFT-SAPT are between CCSD(T) and fixed-node diffusion Monte Carlo (FN-DNC) values.[4] When considering the growth of energy terms in terms of system size, something which has been done already in several instances (e.g., refs 1,73,74), we can already get an idea of the different behavior that correlation methods can exhibit when moving to larger systems.

If we plot interaction energies vs number of atoms (see Supporting Information), we obtain coefficients of determination $R^2$ ranging from 0.79 (MP3) to 0.94 (FN-DMC). The slope of FN-DMC, at 1.09 kJ/mol per atom, is much smaller than that of CCSD(T) at 1.46 kJ/mol per atom — with DFT-SAPT and MP2.5 (1.28 and 1.34 kJ/mol per atom, respectively) both yielding slightly lower slopes than CCSD(T).

**3.2. Intermolecular Interaction Energies of Larger Molecules.** The first question is, whether we can also estimate these numbers from interaction energy slopes of other molecules? If so, they may offer a hint about the accuracy of different post-Hartree−Fock methods for the interaction energies of large molecules. Systems that come to mind are the benzene, coronene, and circumcoronene dimers, as the coronene dimer with 72 atoms was already present in the original L7 set. The parallel-displaced coronene dimer exhibits a significant difference in excess of 10 kJ/mol between FN-DMC, 75.6 kJ/mol, and CCSD(T), 86.1 kJ/mol. As sizable post-CCSD(T) contributions can also be observed for the interaction energies of small π-stacks in general, the question is whether these transfer to larger systems and just "scale up".[26] Given, however, that the circumcoronene dimer already has 144 atoms, we need other, more tractable, systems as touchstones for our hypothesis, hopefully enabling us to infer the behavior for really large systems from the slopes of interaction energies of polyaromatic dimers.

To closely match the aromatic behavior observed in the above sequence, we chose the acene dimer series (displayed in later Figures), thus again starting with the interaction energies of the benzene through hexacene dimers. Starting from heptacene dimer, however, the monomers will undergo cyclodimerization,[75−77] which means that if we aim to relax geometries like in the L7 paper (even with symmetry constraints) we will have to stop at hexacene dimer. The latter, with 82 atoms, is about the same size as coronene dimer. All acenes show very little static correlation, with strong correlation diagnostics close to those of their parent molecule benzene. As a second series, we can mix-and-match species, and arrive at the series including benzene-

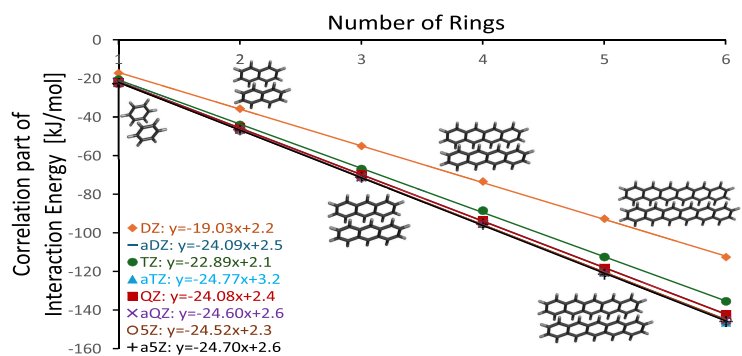

(a) Counterpoise-corrected energies for the acene series.

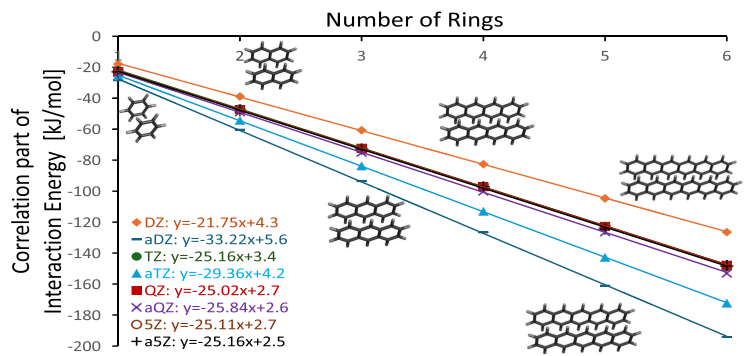

(b) Counterpoise-uncorrected energies for the acene series.

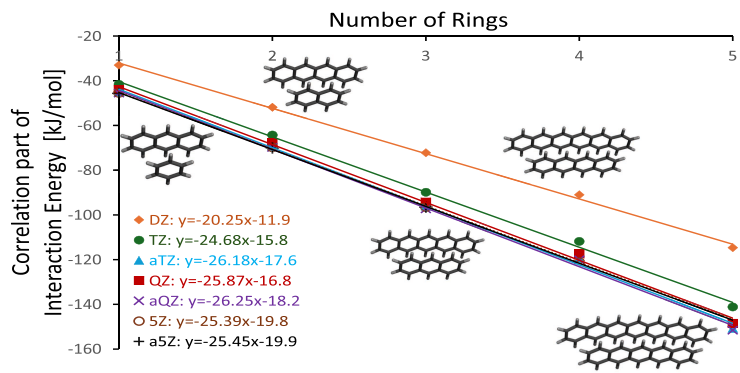

(c) Counterpoise-corrected energies for the acene-(acene-2) series.

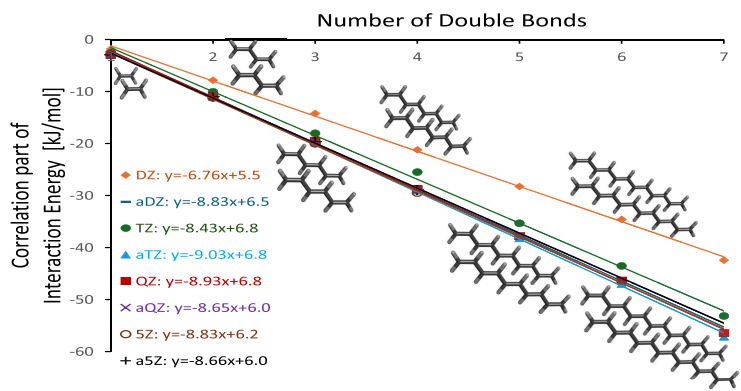

(d) Counterpoise-corrected energies for the polyene stack series (relaxed geometry).

**Figure 1.** Correlation energies in kJ/mol vs number of rings (acene dimers) or number of double bonds (polyene stacks) using the cc-pV$n$Z (X = D − 5) and aug-cc-pV$m$Z ($m$ = D − 5) basis sets.

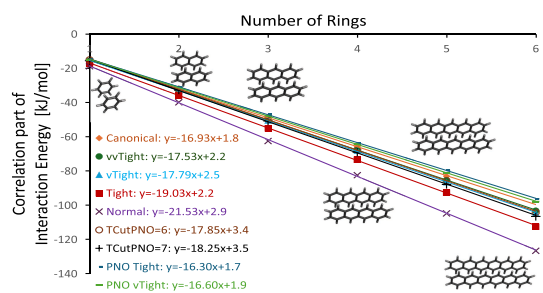

(a) Counterpoise-corrected DZ energies for the acene series.

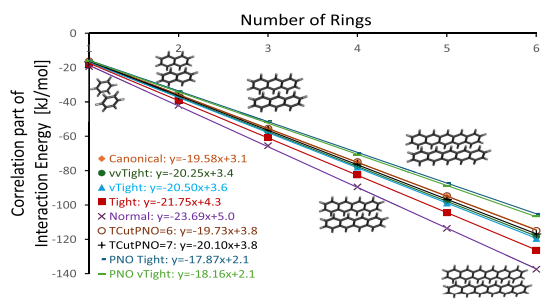

(b) Counterpoise-uncorrected DZ energies for the acene series.

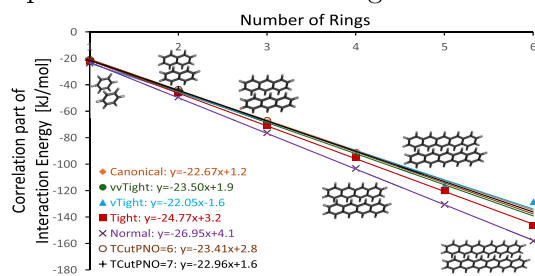

(c) Counterpoise corrected aTZ Energies for the acene series.

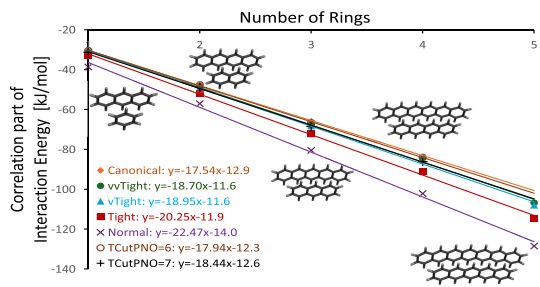

(d) Counterpoise-corrected DZ energies for the acene-acene-2 series.

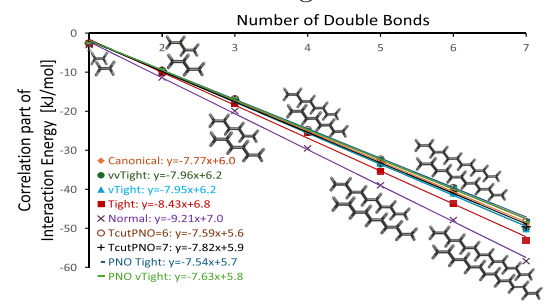

(e) Counterpoise corrected TZ Energies for the polyene stack series (relaxed geometry).

**Figure 2.** Correlation energies (kJ/mol) derived from various coupled cluster methods in as a function of system size: per number of rings (acene dimers) or number of double bonds (polyene stacks).

anthracene, naphthalene-tetracene, anthracene-pentacene, and pentacene-heptacene. In order to retain symmetry of the pi-stacks directly on top of each other, we can construct lines using the $C_NC_N$ and $C_NC_{N-2}$ series (with N denoting the number of rings ranging from one to six/seven).

Finally, for perspective, we will also consider polyene stacks, as these stacks are the possible smallest possible conjugated carbon species which can be investigated, such as the dimers of ethylene, trans-butadiene, hexatriene, octatetraene, and so on. All these species are interesting in the context of quantum chemistry due to their extended $\pi$-conjugation. As the chain length increases, the extent of $\pi$-conjugation also increases, resulting in enhanced stability and stronger interactions between the molecules.

### 3.3. Basis Set Effects.
First, we try to evaluate the basis set effects of the different methods. If, for example, large polarizabilities were a major issue, the dissociation energies along a chain dimer sequence would not be on a straight line, but rather taper off, as the increasing polarization would be increasingly poorly described by a limited number of basis functions. The number of basis functions, however, also increases linearly with system size, yielding almost perfectly straight lines for different acene series, as shown in Figure 1a-c for LNO−CCSD(T) using `tight` settings.

Next, we aim to evaluate the corresponding graphs for the polyene stack series, expecting to obtain results analogous to those observed for the acene stacks.

As becomes apparent, all of the investigated series are on almost perfectly straight lines with an $R^2 > 0.992$ for all basis sets. With increasing basis set size, the lines converge nicely to slopes of $-24.93 \pm 0.23$ kJ per subunit for the acene stacks. Here, the CP-corrected slopes of the aQZ, 5Z, and a5Z basis sets, as well as the non-CP-corrected slopes QZ, aQZ, and a5Z basis sets fall within this range. Interestingly, while the intercepts of these graphs seem highly basis set-dependent, the slopes are much less sensitive to the basis set used. For example, for the CP-corrected correlation energy of the acene series, the intercepts for basis sets of aTZ and a5Z quality deviate by about one-third, but the slopes only by about 3%.

For the polyene stacks, the intercept for TZ deviates from that for a5Z by 13%, but (again) the slope by only 3%. This trend is visible across all systems, suggesting that even quite small basis sets may give an indication about the slope (albeit not the intercept) of the correlation energy and its behavior. When comparing slopes for ethylene vs acene stacks, we find the former to be roughly one-third of the latter. This suggests that there is still some relationship between their behaviors. In contrast, the intercept of the polyene stacks is about three times the intercept of the acene stacks. The polyene stack series has a slope of $-8.76 \pm 0.16$ kJ if we consider all basis sets except DZ. Based on the obtained graphs and considering basis set sizes, we conclude that the QZ basis set is likely the most advantageous for obtaining reliable slopes for both series.

### 3.4. Local Correlation.
Having established the effects of the basis set, we can estimate canonical CCSD(T) at the basis set limit. Furthermore, especially with the advent of local correlation methods and extrapolating in terms of their cutoff parameters,[78−81] we can evaluate their accuracy when going to larger molecules, displayed in Figure 2. For this purpose, we compare these methods, especially for a small basis set of double-$\zeta$ quality. Again, all of the investigated series exhibit nearly perfect linear regression with $R^2$ values of >0.996 for all methods.

When increasing the LNO cutoff settings `Normal` → `Tight` → `VeryTight` → `vvTight` ("very very tight"), the slopes converge toward the canonical value for local CCSD(T). However, there is one caveat: Unlike with basis sets, the differences are not always systematically smaller and in some cases, there is no difference between vtight and vvtight even if the canonical value is not reached. This makes recently promoted complete PNO space (CPS[82]) extrapolations more erratic than, e.g., extrapolations to the complete basis set (CBS) limit. Correlation consistent[58] basis sets are constructed to exhibit smooth and monotonic convergence to the complete basis set limit. In contrast, LNO and DLPNO cutoffs simply define which natural orbitals are neglected- the energies do not have to increase or decrease monotonically. For example, when the DZ basis set is applied, the differences in slopes when going from `Normal` to `Tight` to `VeryTight` to `vvTight` LNO cutoffs to canonical CCSD(T) change by 2.5, 1.2, 0.3, and 0.6 kJ per subunit for the DZ basis set and 2.2, 2.7, -1.5, 0.8 kJ per subunit for the aTZ basis set, respectively. For the latter basis set, we thus obtain a gap from the `VeryTight` to the `vvTight` setting and canonical CCSD(T). From these numbers, if we would just consider LNO−CCSD(T), we would be bound to believe that the slopes are close to converging, as we would get a slope change of 2.2, 2.7, and -1.5 kJ per subunit. This is, however, not the case, as the difference to the canonical slope remains no less than 0.6 kJ per subunit.

This indicates that the CPS extrapolation, especially like in our context, should be investigated further for larger molecules, and its convergence may not be as strict as desirable. DLPNO, while generally rather close to LNO−CCSD(T) with `VeryTight` and `vvTight` cutoffs, sometimes even moves into the wrong direction — *increasing* its slope further as the PNO-cutoff is tightened. Many of these effects appear to come from error compensation effects at looser cutoff values, which however seem to become smaller when going to tighter cutoff values and larger basis sets. Interestingly, this seems to be rather basis set-dependent, with the basis sets without diffuse functions (TZ and QZ) behaving more like the DZ basis set. For these, an extrapolation does not work — while for the aTZ basis set, the CPS extrapolation is close to the canonical CCSD(T) value. Although the differences in slopes when going from `vvTight` LNO−CCSD(T) to canonical CCSD(T) are somewhat larger than e.g., from an aTZ basis set to the CBS limit, these differences are relatively small. Even for slopes of larger molecules, the errors are still mild when using LNO−CCSD(T) and/or a basis set of limited size.

For the polyene stack series, the differences in slopes when going from `Normal` to `Tight` to `VeryTight` to `vvTight` to canonical CCSD(T) change by 0.78, 0.48, −0.01, and 0.19 kJ per subunit, which also implies that the CPS extrapolation would only work for rather loose convergence criteria. In this series, it is noteworthy that the DLPNO−CCSD(T) method with a `TCutPNO` value of $10^{-7}E_h$ slightly underestimates the slope but yields results that are closest to those obtained with the canonical CCSD(T) method. In contrast, the DLPNO−CCSD(T) method with `TCutPNO` set to $10^{-6}E_h$ overestimates the slope relative to the canonical CCSD(T), yet it still provides a more accurate slope estimate than any of the LNO methods. Furthermore, the application of CPS extrapolation produces values that are very close to those obtained with the LNO−CCSD(T) method using a `vvTight` threshold. However, it does not improve upon the accuracy achieved with the `TCutPNO` setting of $10^{-7}E_h$.

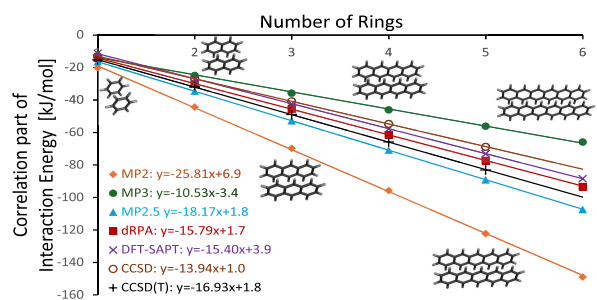

(a) Counterpoise-corrected DZ energies for the acene series.

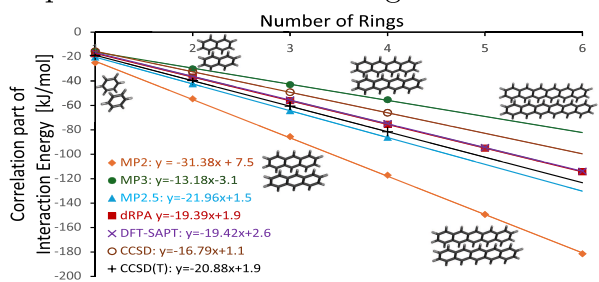

(b) Counterpoise-corrected TZ energies for the acene series.

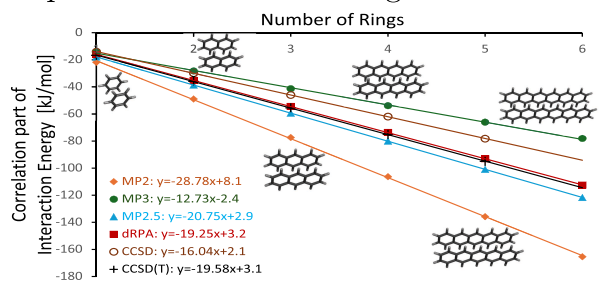

(c) Counterpoise-uncorrected DZ energies for the acene series.

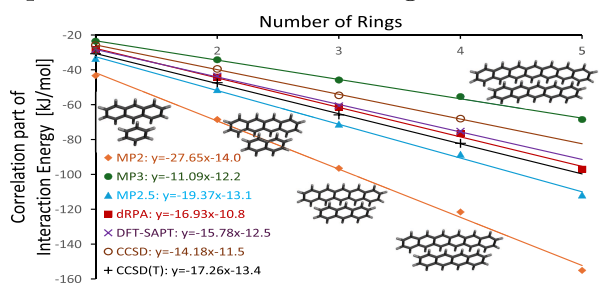

(d) Counterpoise-corrected DZ energies for the acene-acene-2 series.

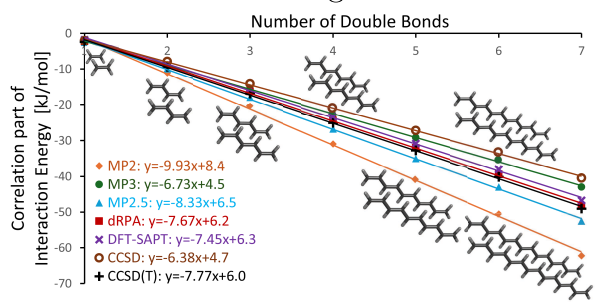

(e) Counterpoise-corrected TZ Energies for the polyene stack series (relaxed geometry).

**Figure 3.** Correlation energies derived from approximate methods in kJ/mol vs number of rings (acene dimers) or number of double bonds (polyene stacks).

Table 2. Best Available CP-Corrected Correlation Energy Slopes (kJ/mol per Ring), Including CCSDT(Q) Values, of the Acene Dimers[a]

| | | method | | | | | | |
|---|---|---|---|---|---|---|---|---|
| | | MP2 | CCSD | CCSD(T) | | | post-CCSD(T) | |
| slopes | basis set | MP2 | CCSD | Tight | VeryTight | canonical | CCSDT-2 | CCSDT | CCSDT(Q) |
| 1 → 2 | DZ | −24.19 | −13.84 | −18.46 | −17.24 | −16.67 | −16.17 | −15.92 | −16.32 |
| 1 → 5 | DZ | −25.56 | −13.94 | −18.86 | −17.67 | −16.93 | −16.43 | −16.18 | −16.58 |
| 1 → 2 | {T,Q}Z | −32.43 | −18.46 | −23.53 | −23.61 | −23.14 | −22.44[b] | −22.09[b] | −22.65[b] |
| 1 → 5 | {T,Q}Z | −33.91 | −18.72 | −24.94 | −24.11 | −23.40 | −22.71[b] | −22.35[b] | −22.97[b] |
| 1 → 6 | {T,Q}Z | −33.97 | −18.78 | −24.96 | | −23.42 | −22.73[b] | −22.37[b] | −22.99[b] |
| 1 → 2 | a{Q,5}Z | | | −23.57 | −23.19 | −22.80 | −22.10[b] | −21.75[b] | −22.31[b] |
| 1 → 3 | a{Q,5}Z | | | −23.95 | −23.49 | −23.10 | −22.40[b] | −22.05[b] | −22.61[b] |
| 1 → 6 | a{Q,5}Z | −33.58 | −18.39 | −24.69 | −24.23 | −23.15 | −22.46[b] | −22.10[b] | −22.72[b] |

[a]Numbers in italics are estimates. [b]The post-CCSD(T) slopes have been scaled by a factor of 1.4 to account for the small cc-pVDZ basis set size; see discussion below and the Supporting Information.

Table 3. Best Available Correlation Energy Slopes (kJ/mol per Ring), Including CCSDT(Q) Values, of the Acene Dimers[a]

| | | method | | | | | | |
|---|---|---|---|---|---|---|---|---|
| | | MP2 | CCSD | CCSD(T) | | | post-CCSD(T) | |
| slopes | basis set | MP2 | CCSD | Tight | VeryTight | canonical | CCSDT | CCSDT(Q) |
| 1 → 2 | DZ | −27.16 | −15.93 | −20.95 | −20.12 | −19.32 | −18.54 | −18.94 |
| 1 → 5 | DZ | −28.53 | −16.04 | −21.71 | −20.41 | −19.49 | −18.71 | −19.11 |
| 1 → 2 | {T,Q}Z | −32.66 | −18.40 | −24.38 | −23.60 | −23.09 | −22.00[b] | −22.56[b] |
| 1 → 5 | {T,Q}Z | −33.92 | −18.57 | −24.89 | | −23.26 | −22.17[b] | −22.73[b] |
| 1 → 6 | {T,Q}Z | −33.95 | −18.60 | −24.92 | | −23.29 | −22.20[b] | −22.76[b] |
| 1 → 2 | a{Q,5}Z | | | −24.09 | −23.00 | −22.48 | −21.39[b] | −21.95[b] |
| 1 → 3 | a{Q,5}Z | | | −24.28 | −23.22 | −22.71 | −21.61[b] | −22.17[b] |
| 1 → 6 | a{Q,5}Z | −33.47 | −18.46 | −24.44 | −23.39 | −22.81 | −21.72[b] | −22.28[b] |

[a]Numbers in italics are estimates. [b]The post-CCSD(T) slopes have been scaled by a factor of 1.4 to account for the small cc-pVDZ basis set size; see discussion below and the Supporting Information.

**3.5. Electronic Structure Methods.** Now that we have established the slopes of canonical CCSD(T) at its basis set limit and thus its performance for large molecules, we assess the performance of other, more approximate methods, shown in Figure 3.

Here, we investigate several post-Hartree–Fock methods, such as MP2, MP2.5, MP3, the dispersion contribution to DFT-SAPT, as well as dRPA. Among these methods, DFT-SAPT and dRPA in particular are deemed to be more accurate than perturbation theory for large molecules.[13,14] For the acene series, we obtain the trend of MP3 < CCSD < DFT-SAPT ≈ dRPA < CCSD(T) < MP2.5 < MP2. This is hardly surprising, as MP2 significantly overestimates the interaction energies of large aromatic molecules, while MP3 underestimates them, as does CCSD. Of course, the question arises about the accuracy of CCSD(T), dRPA, and DFT-SAPT.

For the polyene stack series, we observe the trend CCSD < MP3 < DFT-SAPT < dRPA ≈ CCSD(T) < MP2.5 < MP2. As with the acene series, MP2 overestimates dispersion contributions while both CCSD and MP3 underestimate them. dRPA yields results similar to CCSD(T), as observed with the cc-pV$n$Z (X = D, T, Q) basis sets. (We note[83] that RPA is equivalent to a ring-diagram simplification of coupled cluster with all doubles, CCD.)

**3.6. Effects of the Geometry.** So far, we have used geometries for both acenes and polyene stacks in which the distances between the two monomers were fully optimized within the (high) symmetry imposed. To show that the linear behavior is not an artifact, it is worth investigating another geometry with a different intermonomer distance. Here, we chose to investigate the polyene dimer stacks. At the rather large optimized intermonomer distance of the ethylene dimer, as shorter distances (e.g., at the hexatetraene distance) would result in a positive interaction energy for the ethylene dimer. The polyene stack series with relaxed geometry exhibits a slope of −8.72 ± 0.27 kJ per subunit. In contrast, the same series with fixed geometry at the minimum ethylene dimer distance of 4.46 Å shows a much smaller slope of −3.81 ± 0.24 kJ per subunit.

Results obtained using more approximate methods are identical for both types of geometries, with the only exception being that the slope determined by dRPA is closer to the slope obtained by CCSD(T) in the case of fixed geometries.

All methods analyzed in this study demonstrate better convergence of the LNO approaches, compared to canonical CCSD(T), for relaxed geometries, irrespective of the basis set used. This is attributed to the fact that, for these geometries, the system's energy is closer to the true potential energy minimum. This conclusion is supported by the observation that the absolute values of total energies and correlation energies are larger. Similarly, basis set convergence is more consistent for relaxed geometries, as the electron distribution and correlation effects are more accurately aligned with the optimized structure, resulting in more consistent convergence behavior as the basis set is expanded.

**3.7. Best Estimates of Slopes.** Usually, the best estimates in energies are obtained by going to ever larger basis sets and then just extrapolating the energies to the complete basis set (CBS) limit. In our case, however, we can either calculate ever larger systems along the sequences studied with a smaller basis set, or increase the basis set size, thus adding one dimension to the problem. Our results are summarized in Tables 2 and 3. The slopes obtained from just the first two acene dimers, namely benzene$_2$ and naphthalene$_2$, barely differ compared to those of the whole series up to pentacene; the CP-corrected DZ CCSD(T) slope changes by 0.26 kJ per subunit, and the CP-uncorrected DZ CCSD(T) slope even less at 0.17 kJ per subunit. In general, including larger systems in the fits slightly increases the CCSD(T) slopes. From these values — including the cc-pV{T,Q}Z basis set extrapolation, which is still feasible for canonical CCSD(T) for the benzene and naphthalene dimers — we can estimate the canonical CCSD(T)/cbs values for the benzene-pentacene series, which will be

$$E^{\text{canonical}}_{\text{cc−pV(T,Q)Z}}(1 \to 5) \approx E^{\text{canonical}}_{\text{cc−pV(T,Q)Z}}(1 \to 2) + E^{\text{canonical}}_{\text{cc−pVDZ}}(1 \to 5) - E^{\text{canonical}}_{\text{cc−pVDZ}}(1 \to 2) \quad (4)$$

resulting in

$$-23.14 - 16.93 - (-16.67) = -23.40 \text{ kJ per subunit}$$

for the CP-corrected CCSD(T) results and

$$-23.09 - 16.49 - (-19.32) = -23.26 \text{ kJ per subunit}$$

for CP-uncorrected CCSD(T). Finally, these values can then be extended to the hexacene dimer with an aug-cc-pV{Q,5}Z basis set extrapolation by using the LNO−CCSD(T) values, e.g.

Table 4. Best Available CP-Corrected Correlation Energy Slopes (kJ/mol per Double Bond), Including CCSDT(Q) Values, of the Polyene Stack Relaxed Dimers[a]

| | | method | | | | | | |
|---|---|---|---|---|---|---|---|---|
| | | MP2 | CCSD | CCSD(T) | | post-CCSD(T) | | |
| slopes | basis set | | | vvTight | canonical | CCSDT-2 | CCSDT | CCSDT(Q) |
| 1 → 2 | DZ | −6.53 | −4.74 | −5.58 | −5.47 | −5.35 | −5.36 | −5.49 |
| 1 → 3 | DZ | −6.99 | −4.89 | −5.81 | −5.69 | −5.56 | −5.54 | −5.70 |
| 1 → 4 | DZ | −7.44 | −5.08 | −6.07 | −5.95 | −5.81 | −5.81 | −5.97 |
| 1 → 7 | TZ | −9.93 | −6.38 | −7.96 | −7.77 | −7.57[b] | −7.57[b] | −7.79[b] |
| 1 → 2 | {T,Q}Z | −9.53 | −6.55 | −8.11 | −7.97 | −7.81[b] | −7.82[b] | −8.00[b] |
| 1 → 4 | {T,Q}Z | −10.58 | −6.96 | −8.73 | −8.56 | −8.36[b] | −8.37[b] | −8.59[b] |
| 1 → 7 | {T,Q}Z | | | −9.03 | −8.84 | −8.64[b] | −8.64[b] | −8.86[b] |
| 1 → 2 | a{Q,5}Z | −9.54 | −6.55 | −8.04 | −7.98 | −7.82[b] | −7.82[b] | −8.01[b] |
| 1 → 4 | a{Q,5}Z | | | −8.63 | −8.47 | −8.28[b] | −8.26[b] | −8.50[b] |
| 1 → 7 | a{Q,5}Z | | | −8.94 | −8.74 | −8.55[b] | −8.55[b] | −8.77[b] |

[a]Numbers in italics are estimates. [b]The post-CCSD(T) slopes have been scaled by a factor of 1.4 to account for the small cc-pVDZ basis set size; see discussion below and the Supporting Information.

$$E^{\text{canonical}}_{\text{aug−cc−pV(Q,5)Z}}(1 \to 6) \approx E^{\text{canonical}}_{\text{cc−pV(T,Q)Z}}(1 \to 5) + E^{\text{LNO,tight}}_{\text{aug−cc−pV(Q,5)Z}}(1 \to 5) - E^{\text{LNO,tight}}_{\text{cc−pV(T,Q)Z}}(1 \to 5) \quad (5)$$

arriving at

$$-23.40 - 24.69 - (-24.94) = -23.15 \text{ kJ per subunit}$$

for the CP-corrected and at

$$-23.26 - 24.44 - (-24.89) = -22.81 \text{ kJ per subunit}$$

for the CP-uncorrected CCSD(T) estimate.

Of more interest are the post-CCSD(T) slopes. The CCSDT and CCSDT(Q) results for acene dimers (benzene$_2$ and naphthalene$_2$) included in Tables 2 and 3 were obtained using rank-reduced coupled cluster methods described in refs 84−86. In these calculations, the coupled cluster equations are solved within a certain excitation subspace instead of the full space as in the canonical methods. This allows for reducing the costs of the calculations, but introduces an additional variable (size of the excitation subspace) as a parameter in the rank-reduced formalism. Therefore, it is important that the results are stable with respect to the adopted value of this parameter. In the Supporting Information, we provide a detailed analysis of the convergence of the rank-reduced results with respect to the size of the excitation subspace for benzene dimer and naphthalene dimers, showing that the numerical values reported here are sufficiently well converged for the present purposes. The only additional approximation employed in the rank-reduced calculations is the Cholesky decomposition of two-electron integrals. We used the full pivoting variant of this decomposition with a VeryTight threshold ($10^{-6}E_h$) for the diagonal elements. The errors resulting from the decomposition with these settings impact the calculated interaction energies by no more than a few thousandths of a kJ/mol. The CCSDT method reduces the slopes (in absolute value) by a sizable 0.75 (CP-corrected DZ) and 0.78 (standard DZ) kJ per subunit. However, CCSDT(Q) (again, in absolute value) reverts this by 0.4 in both instances, shrinking the CCSD(T) to CCSDT(Q) differences in the slopes to 0.35 (CP-corrected DZ) and 0.38 (uncorrected DZ) kJ per subunit.

(We note in passing that the (Q) contribution found here for benzene dimer, 0.50 kJ/mol, is just over half of the value reported by Karton and Martin,[87] 0.19−0.20 kcal/mol, or 0.78−0.84 kJ/mol. This was obtained from a thermochemical cycle combined with a CCSDT(Q) calculation in an unpolarized cc-pVDZ(p,s) double-ζ basis set, i.e., [3s2p] on C and [2s] on H, and serves as a cautionary tale in this regard — clearly, one neglects the $d$ functions on carbon at one's peril.)

For the Hartree−Fock exchange part, this is rather simple, as all calculations were performed at the largest basis sets and the CP-corrected and CP-uncorrected slopes are 11.95 and 11.96 kJ per subunit, respectively, yielding exchange-correlation slopes of −10.6 ± 0.3 kJ/mol per acene ring dimer for CCSDT(Q). For the largest acene dimer, hexacene, this would make an interaction energy of approximately 63.6 kJ/mol, considering our best estimates. This value, however, is neglecting the intercepts of 3.3 for Hartree−Fock exchange and 1.9 for correlation, arriving at 58.4 ± 1.8 kJ/mol for CCSDT(Q).

Table 5. Best Available Correlation Energy Slopes (kJ/mol per Double Bond), Including CCSDT(Q) Values, of the Polyene Stack Relaxed Dimers[a]

| | | method | | | | | | |
|---|---|---|---|---|---|---|---|---|
| | | MP2 | CCSD | CCSD(T) | | post-CCSD(T) | | |
| slopes | basis set | MP2 | CCSD | vvTight | canonical | CCSDT-2 | CCSDT | CCSDT(Q) |
| 1 → 2 | DZ | −7.03 | −4.87 | −5.93 | −5.72 | −5.59 | −5.59 | −5.72 |
| 1 → 3 | DZ | −7.60 | −5.15 | −6.24 | −6.09 | −5.94 | −5.92 | −6.09 |
| 1 → 4 | DZ | −8.14 | −5.44 | −6.62 | −6.46 | −6.30 | −6.29 | −6.46 |
| 1 → 7 | TZ | −10.71 | −6.94 | −8.62 | −8.46 | −8.24[b] | −8.23[b] | −8.46[b] |
| 1 → 2 | {T,Q}Z | −9.65 | −6.57 | −8.15 | −8.01 | −7.83[b] | −7.82[b] | −8.01[b] |
| 1 → 4 | {T,Q}Z | −10.68 | −6.98 | −8.75 | −8.59 | −8.37[b] | −8.36[b] | −8.59[b] |
| 1 → 7 | {T,Q}Z | | | −9.01 | −8.85 | −8.63[b] | −8.62[b] | −8.86[b] |
| 1 → 2 | a{Q,5}Z | −9.58 | −6.53 | −8.04 | −7.95 | −7.77[b] | −7.77[b] | −7.96[b] |
| 1 → 4 | a{Q,5}Z | | | −8.59 | −8.43 | −8.20[b] | −8.20[b] | −8.43[b] |
| 1 → 7 | a{Q,5}Z | | | −8.85 | −8.69 | −8.47[b] | −8.46[b] | −8.69[b] |

[a]Numbers in italics are estimates. [b]The post-CCSD(T) slopes have been scaled by a factor of 1.4 to account for the small cc-pVDZ basis set size; see discussion below and the Supporting Information.

CCSD(T), for comparison, would yield slightly larger slopes of −11.3 and −10.9 kJ/mol, with an interaction energy of 66.6 kJ/mol for the hexacene dimer. Including the intercepts (3.3 and 2.0 kJ/mol), we arrive at a final CCSD(T) energy of 61.3 ± 1.8 kJ/mol.

An analogous estimation is performed for the polyene stack series in Tables 4 and 5. To extrapolate CCSDT and CCSDT(Q) values, we utilize the difference between CCSDT and CCSD(T) or between CCSDT(Q) and CCSD(T) results obtained with the DZ basis set. The most accurate CCSD(T) calculations were performed using the TZ basis set for the entire polyene stack series (ethylene, tetraene, pentaene up to tetradecaneheptaene) and the QZ basis set for ethylene–decanepentaene. These results were subsequently used to extrapolate values to larger basis sets. Interestingly, canonical CCSD(T) even slightly underestimates the correlation interaction energy slopes relative to CCSDT(Q), showing that the correlation slopes obtained are rather dependent on the systems chosen. Overall, the post-CCSD(T) contributions are also smaller for the polyene stacks, as the CCSDT−CCSD(T) difference amounts to 2.7% of the correlation contribution to the interaction energy. For the acene series, the CCSDT−CCSD(T) difference is more substantial at 4.6% — presumably because of the aromatic rings.

The more economical CCSD(T)$_\Lambda$ method yields qualitatively the same answer as CCSDT-2, but not quantitatively.

### 3.8. Assessment of the Results when Extrapolating to Larger Molecules.
How would this translate to other, similar systems and their intermolecular interactions, such as the coronene dimer from the L7 set of molecules and the buckycatcher? There are several aspects to this when assessing the post-CCSD(T) correlation effects:

1. Estimate using total interaction energies: Clearly, the interaction energy of the buckycatcher of approximately 175 kJ/mol at the CCSD(T) level of theory is much larger than the interaction energy of the hexacene "sandwich" dimer with our best estimate at approximately 61 kJ/mol. The latter, with six rings, is more comparable to the coronene dimer with seven rings, with an interaction energy of 76 kJ/mol (FN-DMC) to 86 kJ/mol (CCSD-(T)). However, the small discrepancy between CCSD(T) and CCSDT(Q) is rather telling, as it is only 1.5 kJ/mol for the hexacene dimer. Even extrapolating the interaction energy with this slope to the buckycatcher, the deviation would only sum up to approximately 4 kJ/mol difference. This is indicating that the deviations between CCSDT-(Q) and CCSD(T) for intermolecular interactions are an order of magnitude smaller than the 44 kJ/mol difference between FN-DMC and CCSD(T) of Table 1.

2. Estimate using the MP2-CCSD difference in interaction energies: In order to get another, possibly better estimate, we do not compare total interaction energies, but rather the MP2−CCSD difference in interaction energies. The MP2−CCSD difference is as large as 98 kJ/mol for the coronene dimer. For the buckycatcher, the MP2−CCSD difference in the interaction energy is even as big as 261 kJ/mol. With these MP2−CCSD interaction energy differences, we arrive at different numbers by extrapolation than before when estimating them to have the same slopes as for the acenes. The MP2−CCSD difference in interaction energies of the slope is 15.3 ± 0.1 kJ/mol per acene ring at the basis set limit, taking CP-corrected and uncorrected values into account. Interestingly, the MP2−CCSD difference is highly basis set-dependent, as for the DZ basis set, the difference in the slope is only 10.8 ± 0.5 kJ/mol per acene ring. The deviations in the slopes between CCSD(T) and CCSDT-(Q) are, in comparison, only 0.36 ± 0.02 kJ/mol, which is a factor of 40 smaller than the MP2-CCSD difference in the interaction energy per ring. Extrapolating this difference again to the buckycatcher, this would amount to a deviation of 5.6 kJ/mol (and 2.3 kJ/mol for the coronene dimer) from CCSD(T).

3. Estimate using the MP2-CCSD difference in interaction energies taking basis set effects into account: Unfortunately, we have CCSDT(Q) values using only a double-$\zeta$ basis set. However, we can compute energies at more approximate levels in a larger basis set, compare, and then scale the CCSDT(Q)-CCSD(T) differences in the slopes by these values. For the acenes, the MP2-CCSD(T) slopes differ by 20% when going from a double-$\zeta$ basis set to an extrapolated cc-pV{T,Q}Z value. In contrast, the CCSD−CCSD(T) slopes differ by 40% for the same basis sets (DZ and cc-pV{T,Q}Z), and the polyene stacks yield similar results. As we can see from Table 2, the CCSDT-2 slopes are rather close to CCSDT(Q), and the CCSDT-2-CCSD(T) slopes behave slightly more like CCSD−

CCSD(T) for a small basis set of triple-$\zeta$ quality, see Supporting Information. Because of this, we assume that, using the double-$\zeta$ basis set for the CCSD(T)-CCSDT-(Q) slope, the difference is underestimated by a maximum of 40%. Hence, we would obtain a deviation from CCSD(T) of 7.9 kJ/mol for the buckycatcher and 3.2 kJ/mol for the coronene dimer.

These values of CCSDT(Q) underestimating CCSD(T) by 8 and 3 kJ/mol are much smaller than the respective discrepancies of 44 and 10 kJ/mol found for the buckycatcher and coronene dimer, intermediate between FN-DMC and CCSD(T). Hence, our estimates are much closer to CCSD(T) than to FN-DMC (when compared to Table 1).

The large deviations in the slopes of CCSD and MP2 in comparison to that of CCSD(T) and CCSDT(Q) also imply that a different perspective when looking at intermolecular interactions is needed, and approximate methods may unfortunately not provide a full picture.

For the L7 systems and the buckycatcher dimer in question, the polyene stacks are not of as much interest, as their post-CCSD(T) effects can be considered as negligible here. The CCSD(T) and CCSDT(Q) slopes are extremely similar, arriving at the CCSDT(Q) line including HF (not just the correlation energy):

$$y = (-3.3 \pm 0.1 \text{ kJ/mol}) \cdot x + 3.7 \pm 0.2 \text{ kJ/mol}$$

per ethylene double bond.

In comparison, the final best CCSDT(Q) lines obtained for the acenes (including HF) are

$$y = (-10.6 \pm 0.3 \text{ kJ/mol}) \cdot x + 5.2 \pm 0.3 \text{ kJ/mol}$$

for the full set of the six acenes. The interaction energies of benzene ($D_{6h}$) and naphthalene ($D_{2h}$) "sandwich" structures are $-6.5 \pm 0.2$ and $-15.5 \pm 0.2$ kJ/mol, respectively: the resulting slope would be slightly different from the one discussed above because the two data points include the Hartree−Fock association energy, which has its own slope. The interaction energies of ethylene and trans-butadiene, within the geometric constraints applied, are $-0.36 \pm 0.03$ and $-2.92 \pm 0.09$ kJ/mol. These small molecules or the line of the series can serve as reference for other methods when testing the increase in interaction energy for such species.

## 4. CONCLUSIONS

In case one would like to progress to ever larger molecules, we propose a new perspective for analyzing intermolecular interactions of molecular dimers. Although local methods already provide a good, alternative avenue for calculating these interactions, we have unfortunately not quite reached the point where we can draw definite conclusions about the accuracy of electronic structure methods such as the CCSD(T) "gold standard".

Here, we use different dimer series exhibiting very linear increases in correlation energy, such as acene and polyene stacks. We were able to show that none of the following factors disrupt this linear behavior: the choice of basis set; the local correlation cutoffs; the electronic structure method; and fixed vs relaxed geometry. This is evident from all figures in this paper.

The effects of these factors (basis set, local correlation, electronic structure method, geometry) on the slopes are however sizable.

Regarding basis sets, cc-pVTZ is already sufficient to obtain acceptable slopes, with a deviation in the slopes of <1.5 kJ/mol per ring for the acene series. For local correlation methods, only the tightest cutoffs (`vvTight` or `TCutPNO` = $10^{-7}E_h$) achieve this accuracy, whereas other cutoffs can deviate quite significantly, by more than 4 kJ/mol per ring.

Finally, we are able to estimate the interaction of large acenes and polyene stacks which are of similar interaction energy size or MP2-CCSD difference than the L7 set of molecules by comparing CCSD(T) to CCSDT(Q) slopes: Here, the deviations are much smaller than anticipated, with the CCSDT(Q)−CCSD(T) difference in the slope of 0.4 kJ/mol per ring. For comparison, this value is smaller than the deviation of canonical CCSD(T) from local CCSD(T) with the tightest cutoffs or medium-size basis sets of (aug-)TZ quality from the basis set limit, showing that these two effects can still pose major challenges.

By converging all these effects on the slopes to <0.2 kJ/mol per ring or double bond, we can estimate deviations of interaction energies from CCSD(T) at the basis set limit for large acene and polyene stacks, and ultimately for very large molecules including those which are having huge polarizabilities.

In the case of the acene dimer series, CCSD(T) exhibits an interaction energy excess of 3.4% compared to CCSDT(Q) — which is sizable, but nowhere near the level that would explain the discrepancies to FN-DMC.

In a very recent post-CCSD(T) study on the S66 benchmark,[26] it was found that CCSD(T) is very close to CCSDT(Q) for most systems surveyed, but that CCSD(T) overbinds all aromatic π stacks in the data set (complexes **24** through **29**, the six possible dimers of benzene, pyridine, and uracil). For these six species (Table 4 in ref 26), the signed mean difference between FN-DMC from ref 4. and CCSD(T) is $-1.3 \pm 0.7$ kJ/mol, the minus sign indicating that FN-DMC is less bound. (While the present paper was being revised, a preprint[88] with updated FN-DMC numbers came online: for the six aromatic π stacks, the same average with a smaller uncertainty is found, $-1.3 \pm 0.3$ kJ/mol.) The corresponding signed averages for CCSDT−CCSD(T) and CCSD(cT)-CCSD(T), which are $-0.9$ and $-1.0$ kJ/mol, respectively, can be said to agree with overlapping uncertainties with original and revised FN-DMC. However, (Q) binds by an average of $+0.5$ kJ/mol, which brings the CCSDT(Q)−CCSD(T) difference to around $-0.4$ kJ/mol, harder to reconcile with especially the revised FN-DMC data. We surmise that for aromatic stacks, while qualitatively post-CCSD(T) calculations point in the same direction as FN-DMC, a quantitative gap remains.

## ■ ASSOCIATED CONTENT

### ⓈSupporting Information

The Supporting Information is available free of charge at https://pubs.acs.org/doi/10.1021/acs.jctc.4c01512.

- Full raw data, including additional discussions, extra methods used, data for further polyaromatic hydrocarbons (PDF)
- Cartesian coordinates (ZIP)

## ■ AUTHOR INFORMATION

### Corresponding Authors

**Jan M. L. Martin** − *Department of Molecular Chemistry and Materials Science, Weizmann Institute of Science, 7610001*

Reḥovot, Israel; orcid.org/0000-0002-0005-5074; Email: gershom@weizmann.ac.il

A. Daniel Boese − *Department of Chemistry, University of Graz, 8010 Graz, Austria;* orcid.org/0000-0001-7388-778X; Email: adrian_daniel.boese@uni-graz.at

**Authors**

Vladimir Fishman − *Department of Molecular Chemistry and Materials Science, Weizmann Institute of Science, 7610001 Reḥovot, Israel;* orcid.org/0009-0004-7570-136X

Michał Lesiuk − *Quantum Chemistry Laboratory, Faculty of Chemistry, University of Warsaw, 02-093 Warsaw, Poland;* orcid.org/0000-0002-7928-4450

Complete contact information is available at:
https://pubs.acs.org/10.1021/acs.jctc.4c01512

**Notes**

The authors declare no competing financial interest.

## ACKNOWLEDGMENTS

This research was supported by an internal grant to JMLM from the Uriel Arnon Memorial Fund for Artificial Intelligence in Materials Research, as well as by a very generous allocation of CPU time on the Weizmann Faculty of Chemistry HPC facility `chemfarm`. The authors thank its administrators, Dr. Mark Vilensy and Andrei Vasilev, for their kind assistance, and Dr. Margarita Shepelenko for critically reading the draft. The authors acknowledge the financial support by the University of Graz. ADB was supported by the Weizmann Institute of Science as a Weston Visiting Scholar. ML acknowledges Poland's high-performance Infrastructure PLGrid (HPC Centers: ACK Cyfronet AGH, PCSS, CI TASK, WCSS) for providing computer facilities and support within computational grant PLG/2023/016599.